\documentclass{PHYEAUTH}
\usepackage{graphicx}
\usepackage{amsmath}
\usepackage{amssymb}

\begin{document}
\begin{frontmatter}

\title{New Insights into the Plateau-Insulator Transition in the Quantum
Hall Regime}
\author[WZI]{L.A. Ponomarenko \thanksref{cor}},
\thanks[cor]{Corresponding
author. \textit{Email}: leonidp@science.uva.nl}
\author[WZI]{D.T.N. de~Lang},
\author[WZI]{A. de~Visser},
\author[GREN]{D. Maude},
\author[MOSC]{B.N. Zvonkov},
\author[MOSC]{R.A. Lunin}
and
\author[ITF]{A.M.M. Pruisken}

\address[WZI]{Van der Waals-Zeeman Institute, University of
Amsterdam, Valckenierstraat 65, 1018 XE Amsterdam, The
Netherlands}
\address[GREN]{Grenoble High Magnetic Field Laboratory, CNRS-MPI, F-38042 Grenoble, France}
\address[MOSC]{Low Temperature Physics Department, Moscow State
University, 119899, Moscow, Russia}
\address[ITF]{Institute for Theoretical Physics, University of
Amsterdam, Valckenierstraat 65, 1018 XE Amsterdam, The
Netherlands}

\begin{abstract}
We have measured the quantum critical behavior of the
plateau-insulator (PI) transition in a low-mobility InGaAs/GaAs
quantum well. The longitudinal resistivity measured for two
different values of the electron density follows an exponential
law, from which we extract critical exponents $\kappa$ = 0.54 and
0.58, in good agreement with the value ($\kappa$ = 0.57)
previously obtained for an InGaAs/InP heterostructure. This
provides evidence for a non-Fermi liquid critical exponent. By
reversing the direction of the magnetic field we find that the
averaged Hall resistance remains quantized at the plateau value
$h/e^2$ through the PI transition. From the deviations of the Hall
resistance from the quantized value, we obtain the corrections to
scaling.
\end{abstract}


\end{frontmatter}

\begin{figure}[htb]
\begin{center}\leavevmode
\includegraphics[width=0.8\linewidth]{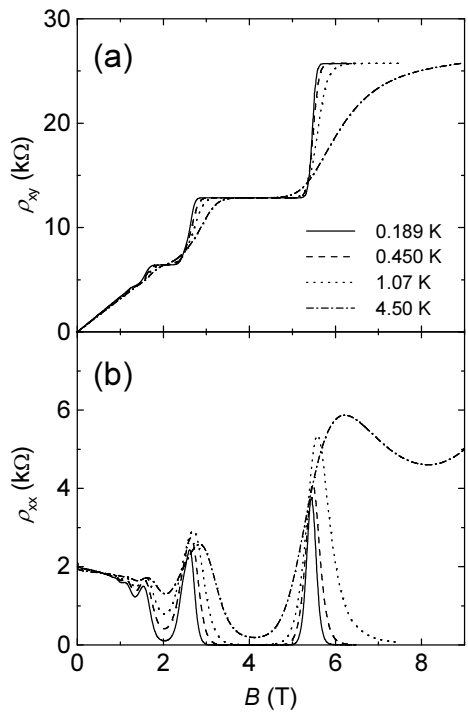}
\caption{The Hall (a) and longitudinal (b) resistivities of
In$_{0.2}$Ga$_{0.8}$As/GaAs quantum well as a function of magnetic
field at selected temperatures.}\label{fig1}\end{center}
\end{figure}
Despite its relatively long history of study, the exact nature of
the transitions between adjacent quantum Hall states is still a
subject of debate. In the framework of the scaling theory of the
quantum Hall effect, the plateau transitions are interpreted as
quantum phase transitions with an associated universal critical
exponent $\kappa  =  p/2\chi$ \cite{pruisken:prl88}. Here $p$ is
the exponent of the phase breaking length $l_\varphi$ at finite
temperature $T$ (i.e. $l_\varphi \sim T^{-p/2}$ ) and $\chi$ the
critical exponent for the zero $T$ localization length $\xi$.
Pioneering magnetotransport experiments by Wei \textit{et
al}.~\cite{wei:prl88} on the plateau-plateau (PP) transitions of a
low mobility InGaAs/InP heterojunction resulted in a value of
$\kappa$ = 0.42. From subsequent current scaling transport
measurements~\cite{wei:prb92} a value $p$ = 2 was extracted.
Hence, an experimental estimate for the localization length
exponent is $\chi \sim 2.4$, which is close to the free electron
value 7/3 obtained from numerical simulations
\cite{huckestein:prl90}. This Fermi liquid result has been quite
puzzling, as a microscopic theory of the quantum Hall effect
should have important ingredients such as localization effects and
Coulomb interactions. It turns out however that the PP transitions
suffer from systematic errors that are inherently due to
macroscopic sample inhomogeneities~
\cite{ponomarenko:condmat,pruisken:condmat}. More recently, we
have conducted magnetotransport experiments~
\cite{pruisken:condmat,schaijk:prl00,lang:pe02} on the
plateau-insulator (PI) transition of an InGaAs/InP
heterostructure. By employing a new methodology, which recognizes
the fundamental symmetries of the quantum transport problem, the
magnetotransport data at the PI transition can be used to separate
the universal critical behavior from the sample dependent
aspects~\cite{pruisken:condmat}. This resulted in a value of the
exponent $\kappa = 0.57$. Since $p$ is bounded by
$1~<~p~<~2$~\cite{baranov:condmat}, this value of $\kappa$ implies
that the correlation length exponent is bounded by $0.9 < \chi <
1.8$, which is in conflict with Fermi liquid ideas.

This important new result asks for experimental verification by
investigating different low-mobility semiconductor structures. The
use of low-mobility structures is dictated by the scattering
mechanism, which should predominantly be short-range random
potential scattering, which results in a wide temperature regime
for scaling. Here we report magnetotransport experiments at the PI
transition of an In$_{0.2}$Ga$_{0.8}$As/GaAs quantum well for two
different values of the electron density $n$.

The In$_{0.2}$Ga$_{0.8}$As/GaAs quantum well (QW) was grown by the
MBE technique. The two-dimensional electron gas is located in a 12
nm thick In$_{0.2}$Ga$_{0.8}$As layer, sandwiched between GaAs.
Carriers are provided by Si $\delta$-doping, separated from the
quantum well by a spacer layer of 20 nm. Hall bars were made with
Au-Ni-Ge contacts. The width of Hall bar equals 75 $\mu$m, and the
distance between the voltage contacts along the Hall bar is 390
$\mu$m. The sample is insulating in the dark state. An infra-red
LED was used to create carriers.

Magnetotransport data up to 9 T were taken in a vacuum loading
Oxford dilution refrigerator. High-magnetic field data up to 20 T
were taken at the Grenoble High Magnetic Field Laboratory using an
Oxford top-loading plastic dilution refrigerator. The longitudinal
$R_{xx}$ and Hall $R_{xy}$ resistances were measured using
standard lock-in techniques at a frequency of 2.1~Hz.
Since the longitudinal resistance exponentially
increases at the PI transition, the phase change of the signal was
carefully monitored. For the data presented here the out-of-phase
signal is always less than 10\%.

In order to characterize our In$_{0.2}$Ga$_{0.8}$As/GaAs QW we
have first measured the low-temperature magnetotransport
properties in fields up to 9 T (see Fig.\ref{fig1}). The sample
was illuminated such that the electron density, as determined from
the initial slope of the Hall resistance at $T$ = 4.5 K, amounts
to $n = 1.9 \times 10^{11}$ cm$^{-2}$. The transport mobility
$\mu$ = 16000 cm$^2/$Vs. The spin resolved $\nu = 2 \rightarrow 1$
PP transition takes place around 5.4 T and is well defined below
$\sim$1.5 K. The presence of weak macroscopic sample
inhomogeneities is illustrated by a small (1\%) carrier density
gradient along the Hall bar (i.e. between the two pairs of Hall
contacts)~\cite{ponomarenko:condmat}. The $R_{xy}(B)$ and
$R_{xx}(B)$ curves of the $\nu = 2 \rightarrow 1$ PP transition,
measured at low $T$, do not show a proper crossing point, while
the peak height $R_{xx}(B)$ remains temperature dependent. The
analysis of the PP transition requires higher-order correction
terms in the transport problem, and will be reported elsewhere.

\begin{figure}[htb]
\begin{center}\leavevmode
\includegraphics[width=0.8\linewidth]{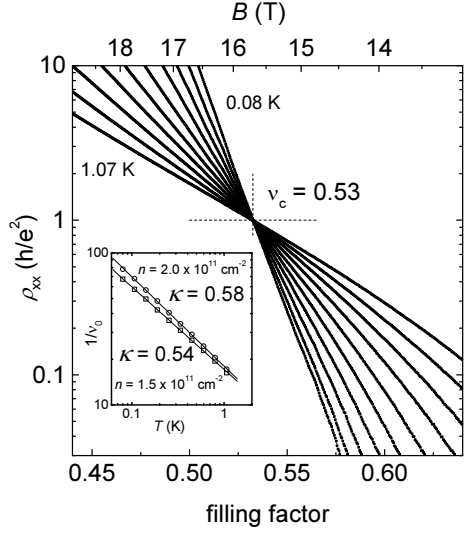}
\caption{The longitudinal resistivity $\rho_{xx}$ as a function of
filling factor $\nu$ at temperatures 1.07, 0.80, 0.60, 0.45, 0.34,
0.26, 0.19, 0.142, 0.107 and 0.080 K for the InGaAs/GaAs QW. The
crossing point indicates the PI transition. Near the critical
point the resistivity $\rho_{xx}$ obeys Eq.(1). The carrier
density $n = 2.0 \times 10^{11}$~cm$^{-2}$. Inset: temperature
dependence of the fitting parameter $1/\nu_0$ (open symbols) for
two different values of carrier densities as indicated. Solid
lines represent linear fits. }\label{fig2}\end{center}
\end{figure}
Low-temperature high-field magnetotransport data near the PI
transition were taken for two different carrier densities, $n =
1.5 \times 10^{11}$ cm$^{-2}$ and $2.0 \times 10^{11}$~cm$^{-2}$.
For these densities the PI transition takes place at $B_c$ = 10.7
and 15.7 T, respectively. Typical results are shown in
Fig.~\ref{fig2}, where we have plotted $\rho_{xx}$ in units
$h/e^2$ on a logarithmic scale versus the filling factor $\nu$.
The data, taken in the temperature range 0.08-1.07 K, show a sharp
crossing point at the critical filling factor $\nu_c =$  0.58 and
0.53 for lower and higher densities, respectively. At the critical
point $\rho_{xx,c} = h/e^2$ (to within 1 \%), as it should. In the
vicinity of $B_c$, the longitudinal resistance $\rho_{xx}$ follows
the empirical law
\begin{equation}
\ln(\rho_{xx}/\rho_0)= -\Delta \nu / \nu_0(T)
\label{phenomenological}
\end{equation}
where $\rho_0 = \rho_{xx,c}$ and $\Delta \nu = \nu - \nu_c$. The
slope $1/\nu_0$ of the curves in Fig.~\ref{fig2} has been
determined by a fitting procedure and obeys power law behavior
($\nu_0 \propto T^\kappa$) over more than one decade in $T$. From
the log-log plot of $\nu_0^{-1}$ vs $T$ we extract a critical
exponent $\kappa = 0.54 \pm 0.02$ and $0.58 \pm 0.02$,
respectively.
\begin{figure}[htb]
\begin{center}\leavevmode
\includegraphics[width=0.8\linewidth]{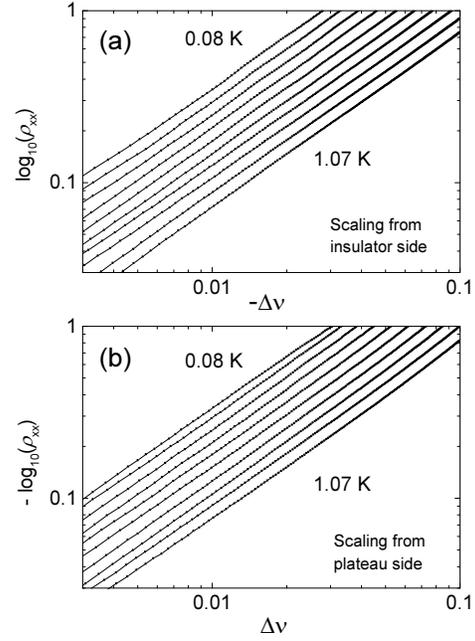}
\caption{The $\rho_{xx}$ data for the InGaAs/GaAs QW ($B_c$ =
15.7~T) from Fig.~\ref{fig2}, plotted versus $\Delta \nu$ in the
insulating (a) and quantum Hall phase (b). The axis are rescaled
to illustrate the validity of Eq.~\ref{eq3}. Equally spaced
parallel lines signify scaling.} \label{fig3}\end{center}
\end{figure}

In order to corroborate this result further, we present in this
paragraph an alternative way of investigating scaling at the PI
transition. Notice that we have chosen the temperatures $T_i$ of
the field sweeps such that these are equally spaced on a
logarithmic temperature scale, i.e. $\ln T_i = \ln T_0 - \alpha
i$, where $T_0$ and $\alpha$ are constants and $i$ is an integer
which labels the different curves. In the case of scaling $\ln
\nu_0(T) = \kappa \ln T$ . Hence, the parameter $\nu_0$ changes
with index $i$ as
\begin{equation}
\ln \nu_0(T_i) = - \alpha \kappa i + const. \label{nuvsT}
\end{equation}
Taking the logarithm of Eq.(\ref{phenomenological}) and taking
into account Eq.(\ref{nuvsT}), we obtain
\begin{equation}
\ln | \ln \rho_{xx} | = \ln | - \Delta \nu | + \alpha \kappa i -
const, \label{eq3}
\end{equation}
\begin{figure}[htb]
\begin{center}\leavevmode
\includegraphics[width=0.8\linewidth]{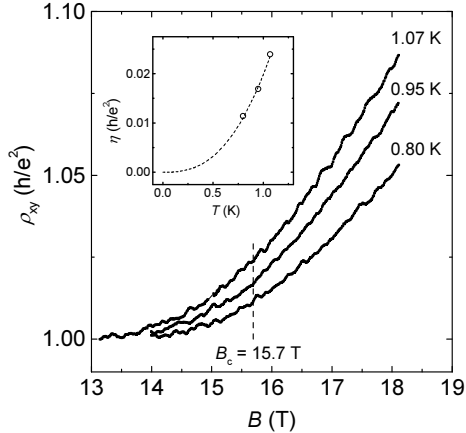}
\caption{The field dependence of the Hall resistivity $\rho_H(B)$
near the critical field $B_c$ at three different temperatures as
indicated. Inset: the deviation of $\rho_H$ from the quantized
value $h/e^2$  at critical field $B_c$ versus $T$.}
\label{fig4}\end{center}
\end{figure}
where $\rho_{xx}$ is expressed in units of $h/e^2$. The absolute
value here is used to take into account both positive and negative
values of $\Delta \nu$ and $\ln \rho_{xx}$, depending whether $B <
B_c$ or $B > B_c$. Under the condition of scaling, a plot of $\ln
| \ln \rho_{xx} |$ vs $\ln | - \Delta \nu |$  transforms the
experimental data into two sets of parallel lines, equally spaced
by the amount $\alpha \kappa$ along the abscissa. This is
illustrated in Fig.~\ref{fig3} for the data set shown in
Fig.~\ref{fig2} ($B_c$ = 15.7 T). The derived values of $\kappa$
are identical to the values quoted above. The advantage of this
method, which relies on the presence of a sharp well-defined
crossing point, over the ``traditional'' method is that no fitting
procedure is needed to visualize scaling behavior. Hence, errors
in the determination of the critical exponent due to arbitrary
fitting constraints are minimized. Besides, scaling behavior can
be verified away from the critical field $B_c$. Furthermore, this
method allows for a direct check of particle-hole symmetry,
$\rho_{xx}(\Delta \nu) = 1/ \rho_{xx}(-\Delta \nu)$. In this case,
the curves for $B > B_c$ (Fig.~\ref{fig3}a) and $B < B_c$
(Fig.~\ref{fig3}b) should be identical as can be verified by
plotting the data from both panels in the same graph.

By reversing the magnetic field polarity, the $\rho_{xx}$ data for
the PI transition remain identical. This is an important
experimental result, as it indicates that the admixture of
$\rho_{xy}$ into $\rho_{xx}$ is negligible, which simplifies the
analysis~\cite{pruisken:condmat}. Thus a proper critical exponent
can be derived directly from the symmetric (with respect to the
magnetic field) $\rho_{xx}$ data. Notice that this is not the case
for the PP transition, where
the $\rho_{xx}$ data for fields up and down are slightly
different.

On the other hand, the Hall resistance data are strongly affected
by reversal of the magnetic field polarity. After averaging over
both field polarities we find that the antisymmetric part of the
Hall resistance, $\rho_H$, remains quantized at the value $h/e^2$
in the insulating state for $T \leq$ 0.2 K. With increasing
temperature, however, $\rho_H(B)$ starts to deviate from the
quantized value. In Fig.~\ref{fig4} we show the averaged Hall
resistance for a density $2.0 \times 10^{11}$ cm$^{-2}$ ($B_c$ =
15.7 T) in the field range 13-18 T at three different
temperatures. At $B_c$, the deviation from exact quantization can
be written as $\rho_H = 1 + \eta(T)$, where $\eta(T) =
(T/T_1)^{y_{\sigma}}$  contains the corrections to scaling. In the
inset of Fig.~\ref{fig4} we have plotted $\eta(T)$ (dashed line)
from which we extract parameters $y_\sigma$ = 2.6 and $T_1$ = 4.5
K, which are similar to the values reported earlier for an
InGaAs/InP heterostructure~\cite{pruisken:condmat}.

In summary, we have studied the quantum critical behavior of the
PI
transition in a low-mobility InGaAs/GaAs quantum well for two
different values of the electron density. From the $\rho_{xx}$
data we extract critical exponents $\kappa$ = 0.54 and 0.58, in
good agreement with the value ($\kappa$ = 0.57) previously
obtained for an InGaAs/InP heterostructure. This provides further
evidence for a non-Fermi liquid quantum critical point. By
reversing the direction of the magnetic field, we find that the
averaged Hall resistance remains quantized at the plateau value
$h/e^2$ through the PI transition. From the deviations of the Hall
resistance from the quantized value, we obtain the corrections to
scaling.
\\

\textit{Acknowledgements.} We are grateful to P.~Nouwens,
V.~A.~Kulbachinskii and Yu.~A.~Danilov for assistance in various
stages of the experiments. This work was supported by the Dutch
Foundation for Fundamental Research of Matter (FOM) and RFBR
01-02-16441.

\end{document}